\documentclass[aps,prl,a4paper,twocolumn,showpacs]{revtex4}
\usepackage[dvips]{epsfig}
\begin{document}
\title{Three-body non-additive forces between spin-polarized alkali atoms}
\author{Pavel Sold\'{a}n} 
\author{Marko T. Cvita\v s} 
\author{Jeremy M. Hutson}
\affiliation{Department of Chemistry, University of Durham, South Road,
Durham, DH1~3LE, England}
\date{\today}

\begin{abstract}
Three-body non-additive forces in systems of three spin-polarized
alkali atoms (Li, Na, K, Rb and Cs) are investigated using
high-level {\em ab initio} calculations. The non-additive forces
are found to be large, especially near the equilateral equilibrium
geometries. For Li, they increase the three-atom potential well
depth by a factor of 4 and reduce the equilibrium interatomic
distance by 0.9 \AA. The non-additive forces originate principally
from chemical bonding arising from sp mixing effects.

\end{abstract}
\pacs{34.20.Mq,31.50.Bc}

\maketitle

\font\smallfont=cmr7

There is at present great interest in the properties of
Bose-Einstein condensates (BECs) formed in dilute gases of alkali
atoms, and in particular in molecule formation in condensates by
processes such as photoassociation \cite{Wyn00,Ger00,McK02},
magnetic tuning of Feshbach resonances \cite{Don02} and three-body
recombination \cite{ketterle98,wieman00}. Once molecules have been formed,
their fate depends largely on collisional processes such as
inelastic and reactive scattering, which can release kinetic
energy and result in the molecules being ejected from the trap.
Calculations on such processes require potential energy surfaces
for the three-atom system. In the absence of better information,
calculations on three-body recombination have mostly used
pairwise-additive three-atom potentials based on atom-atom pair
potentials \cite{greene}. The rationale for this is that
spin-polarized alkali atoms are ``honorary rare gas atoms'', and
that the binding between them is dominated by dispersion forces,
which are nearly pairwise additive. However, it is known that
non-additive forces are significant in spin-polarized Na$_3$
\cite{Hig00}, and we have recently shown \cite{Sol02} that such
terms can affect ultra-low-energy cross sections for the process
Na + Na$_2(v=1) \rightarrow {\rm Na} + {\rm Na}_2(v=0)$ by at
least a factor of 10.

The purpose of the present paper is to investigate the magnitude
of the nonadditive terms for the complete series of homonuclear
alkali trimers. We focus on potential energy surfaces for quartet
states of the three-atom systems, corresponding to interaction of
spin-polarized atoms. These are the surfaces that are most
important in condensates, and are also the ones for which pairwise
additivity appears to be a sensible first approximation. The
corresponding doublet surfaces involve strong chemical bonding,
and are complicated by the presence of conical intersections at
equilateral geometries \cite{doublets1,doublets2}.

The approach we have taken is to perform {\em ab initio}
calculations using a single-reference restricted open-shell
variant \cite{KHW93} of the coupled cluster method \cite{Cizek}
with single, double and noniterative triple excitations
[RCCSD(T)]. All the calculations were performed using the MOLPRO
package \cite{MOLPRO}. The three-atom interaction potential can be
decomposed into a sum of additive and non-additive contributions,
\begin{equation}
V_{\rm trimer}(r_{12},r_{23},r_{13}) = \sum_{i<j} V_{\rm
dimer}(r_{ij}) + V_{3}(r_{12},r_{23},r_{13}). \label{eq2}
\end{equation}
The full counterpoise correction of Boys and Bernardi \cite{BSSE}
was employed to compensate for basis set superposition error in
both dimer and trimer calculations.

The basis sets used in the {\it ab initio} calculations were as
follows. 
For Li  and Na we used the same basis sets as were used by Halls 
{\it et al.} \cite{Li1} and by Gutowski and coworkers 
\cite{Hig00,Gut99} respectively.
For K, Rb and Cs, we used
the small-core ECP10MWB, ECP28MWB and ECP46MWB effective core
potentials (ECPs) of Leininger {\it et al.} \cite{ECP}. These
quasirelativistic \cite{WB} ECPs treat the 1s$^{2}\cdots
$($n$-2)s$^{2}$($n$-2)p$^{6}$($n$-2)d$^{10}$ electrons as core and
the ($n$-1)s$^{2} (n$-1)p$^{6} n$s$^{1}$ electrons as valence. To these
ECPs, medium-size uncontracted valence basis sets were added
\cite{Kbas,Rbbas,Csbas}. The resulting atomic electric dipole
polarizabilities for Li, K, Rb and Cs 
(165.6, 294.2, 319.34 and 402.38
$a_0^3$) are in excellent agreement with the corresponding
experimental values \cite{Polar1,Polar2} 
($164.0 \pm 3.4$, $292.8 \pm 6.1$, $319.2 \pm 6.1$
and $402.2 \pm 8.1 \ a_0^3$), while that for Na is slightly too high
(166.3 $a_0^3$ compared to $162.7 \pm 0.8\ a_0^3$).

\begin{figure}[b]
\begin{center}
\epsfig{file=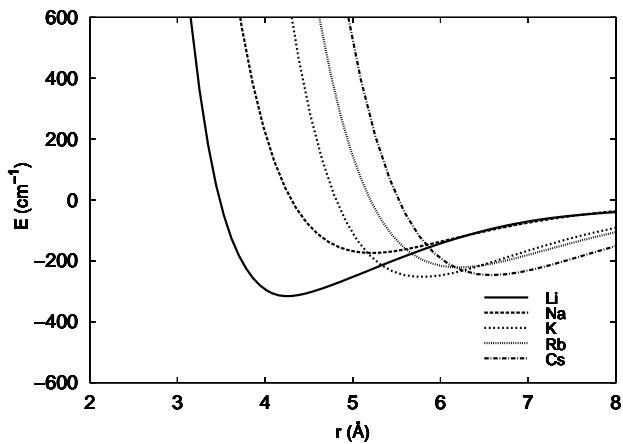,angle=-90,width=85mm}
\end{center}
\caption{RCCSD(T) interaction energies of spin-polarized alkali
dimers.} \label{fig1}
\end{figure}

\begin{figure}[t]
\begin{center}
\epsfig{file=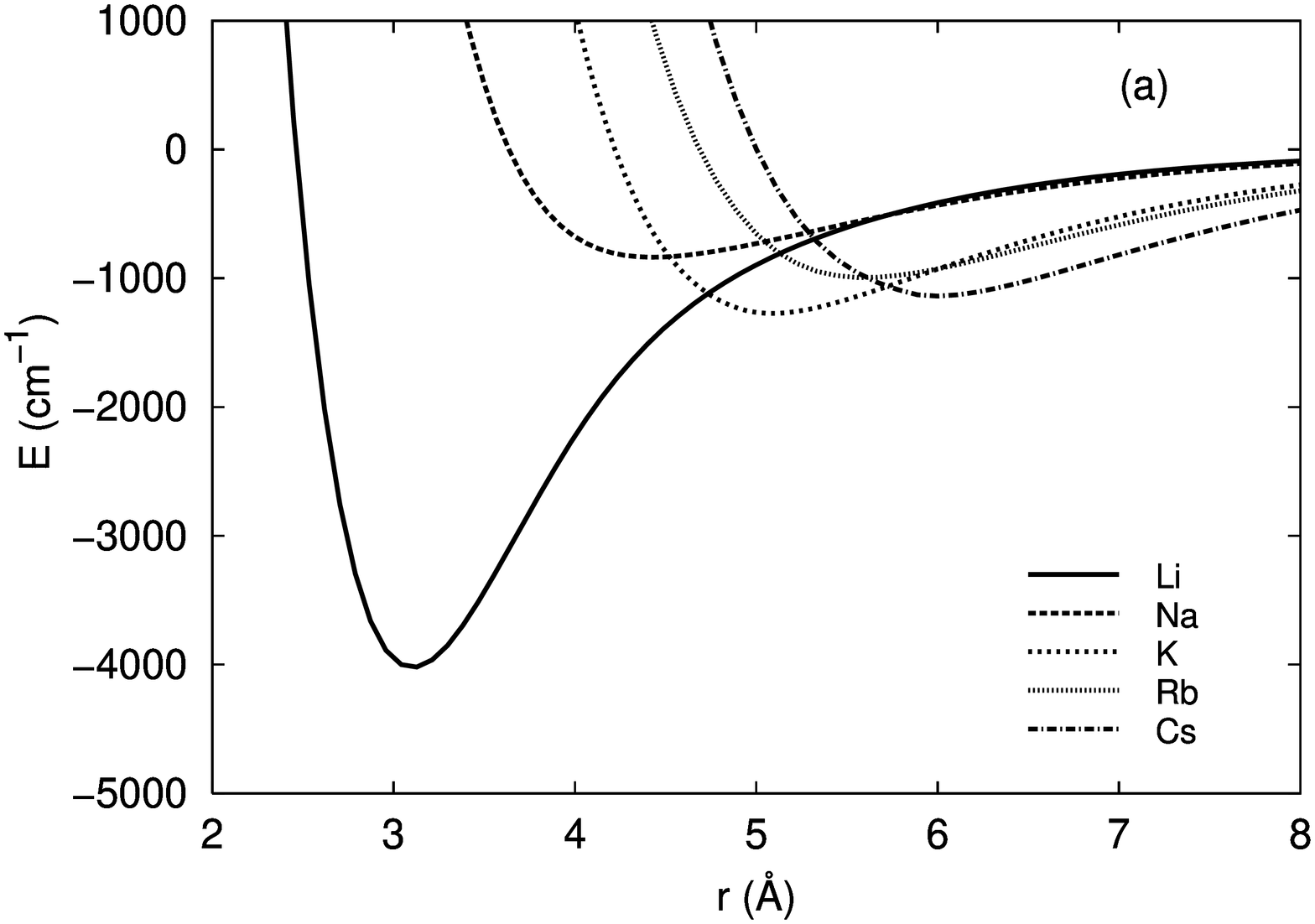,angle=-90,width=85mm}
\epsfig{file=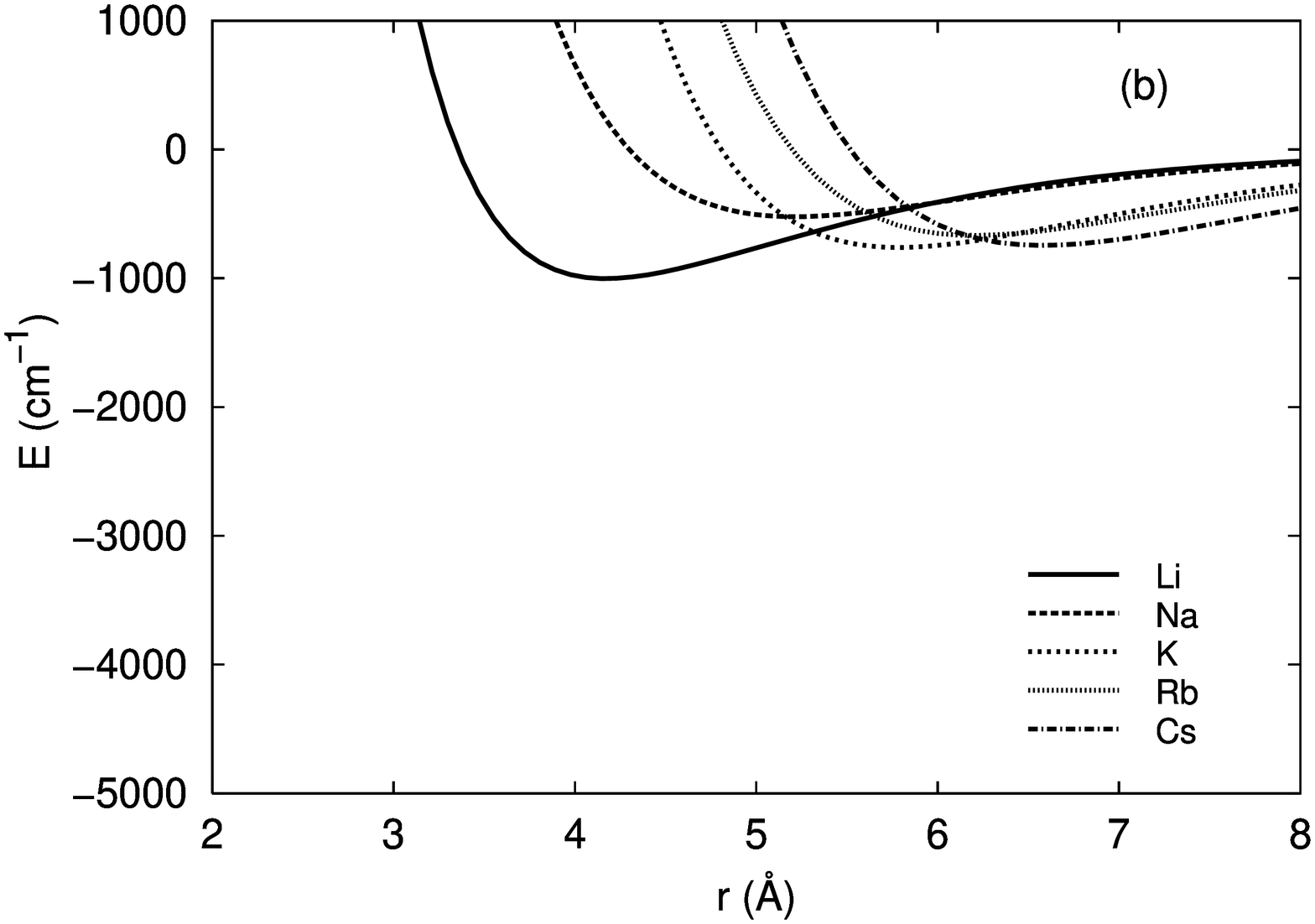,angle=-90,width=85mm}
\end{center}
\caption{RCCSD(T) interaction energies of spin-polarized alkali
trimers at $D_{3h}$ geometries (a) total nonadditive potentials; 
(b) additive potentials. } \label{fig2}
\end{figure}

The RCCSD(T) potential energy curves obtained for the a
$^3\Sigma^+_u$ states of the alkali dimers are shown in Fig.\
\ref{fig1}. The curve characteristics, $V_{\rm min}=-D_{\rm e}$ 
and $r_{\rm e}$, are summarized in Table \ref{table1}. The RCCSD(T) 
results for Li$_2$ agree very closely  (within 0.1 cm$^{-1}$)
with those obtained by Halls {\em et al.}\ \cite{Li1} using
unrestricted quadratic configuration interaction calculations
($D_{\rm e} = 334.145$ cm$^{-1}$, $r_{\rm e} = 4.1686$ \AA), 
which in turn give excellent agreement with the RKR curve obtained
from optical-optical double resonance (OODR) spectra on $^7$Li$_2$
($D_{\rm e} = 333.76 \pm 0.02$ cm$^{-1}$, $r_{\rm e} = 4.173$ \AA)
\cite{Li2,Li3}. The RCCSD(T) results for Na$_{2}$ 
agree very well with those obtained by Gutowski \cite{Gut99} 
using unrestricted coupled-cluster calculations and employing 
the same basis set 
($D_{\rm e} = 173.926$ cm$^{-1}$, $r_{\rm e} = 5.218$ \AA). 
Neither our results nor Gutowski's agree very well with the RKR curve 
obtained from OODR spectra ($D_{\rm e} = 175.76 \pm 0.35$ cm$^{-1}$, 
$r_{\rm e} = 5.108(5)$ \AA) \cite{Na1}, but the
accuracy of the RKR curve for Na$_2$ has been questioned \cite{Gut99,Iva01}.
For K$_2$, the RCCSD(T) results 
are in very good agreement with experimental results on $^{39}$K$_{2}$ 
($D_{\rm e} = 252.74 \pm 0.12$ cm$^{-1}$, $r_{\rm e} = 5.7725(20)$ \AA) 
\cite{K1,K2}.

\begin{figure}[t]
\begin{center}
\epsfig{file=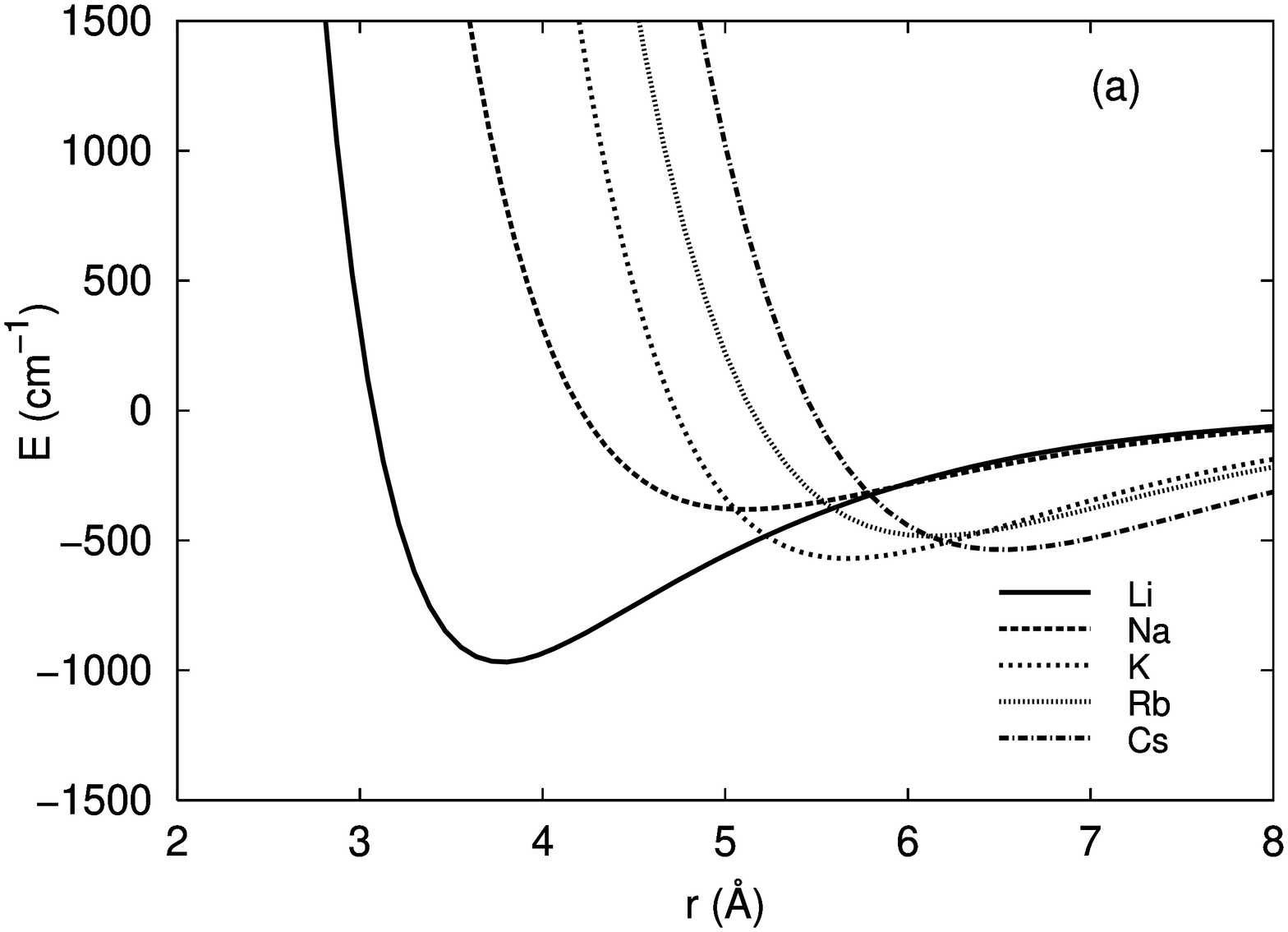,angle=-90,width=85mm}
\epsfig{file=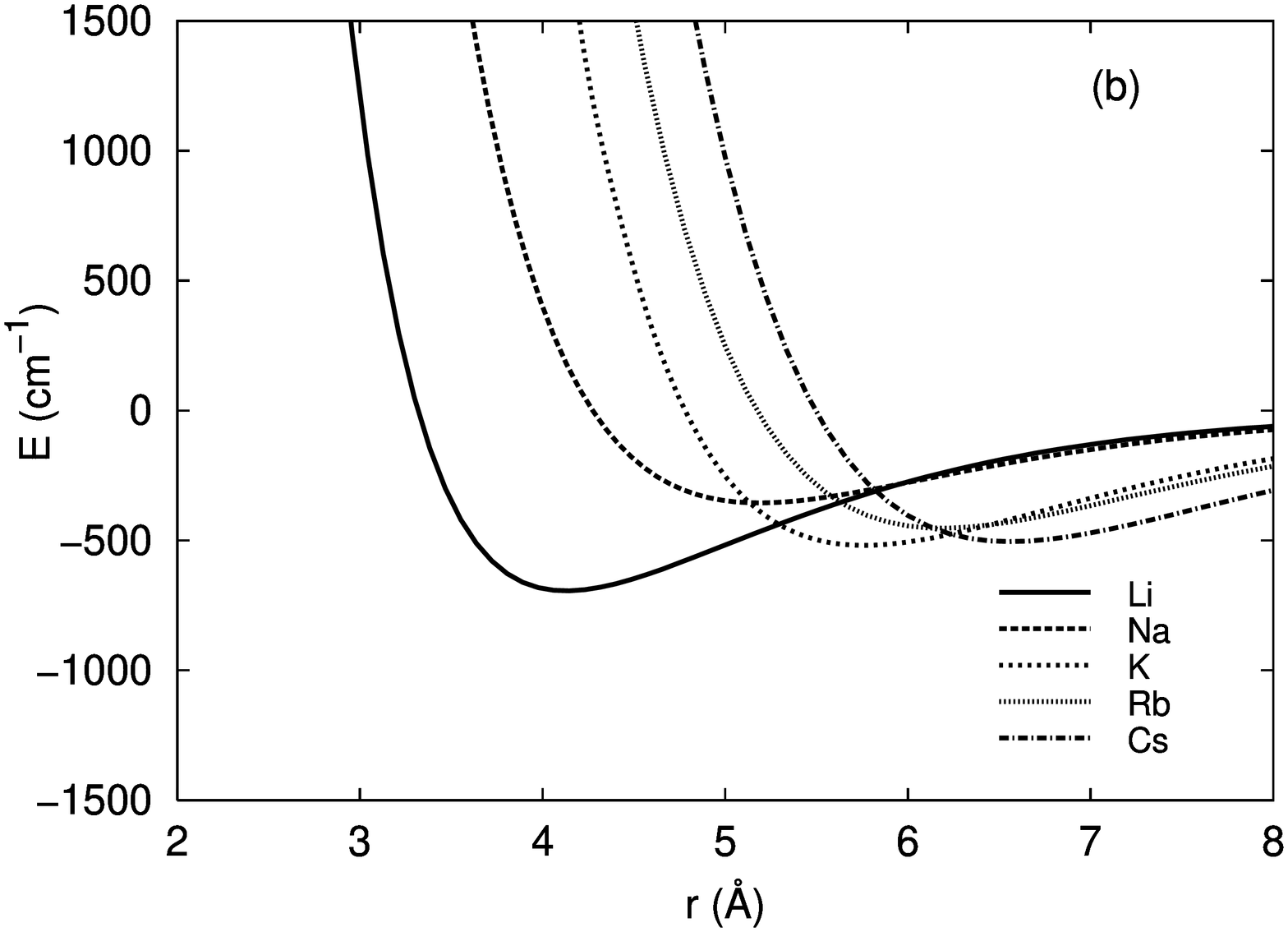,angle=-90,width=85mm}
\end{center}
\caption{RCCSD(T) interaction energies of spin-polarized alkali
trimers at the $D_{\infty h}$ geometries (a) total nonadditive
potentials; (b) additive potentials. }
\label{fig3}
\end{figure}

The RCCSD(T) potentials obtained for the ground states of the
quartet alkali trimers are shown in Figs.\ \ref{fig2} and
\ref{fig3} for equilateral ($D_{3h}$) and symmetric linear
($D_{\infty h}$) geometries respectively. The full trimer
potentials are compared with pairwise-additive potentials based on
the triplet dimer potential in each case. The potential characteristics,
$V_{\rm min}=-D_{\rm e}$ and $r_{\rm e}$ for the global minima and 
$V_{\rm sp}$ and $r_{\rm sp}$ for the linear saddle points, are listed in 
Table \ref{table1}.

\begin{table}[t]
\caption{RCCSD(T) values of $r_{\rm e}$ (\AA), $r_{\rm sp}$ (\AA), 
$V_{\rm min}=-D_{\rm e}$ (cm$^{-1}$), $V_{\rm sp}$ (cm$^{-1}$), 
and $V_3$ (cm$^{-1}$) for spin-polarized alkali dimers and trimers.} 
\label{table1}
\begin{ruledtabular}
\begin{tabular}{lrrrrrrrr}
 & \multicolumn{2}{c}{Dimer} & \multicolumn{3}{c}{Trimer $D_{3h}$} & 
\multicolumn{3}{c}{Trimer $D_{\infty h}$} \\
\cline{2-3} \cline{4-6} \cline{7-9}
 & $r_{\rm e}$ & $V_{\rm min}$ & $r_{\rm e}$ & $V_{\rm min}$ &  $V_{\rm 3}$ & $r_{\rm sp}$ & $V_{\rm sp}$ &  $V_{3}$ \\
\tableline
Li & 4.169 & -334.046 & 3.103 & -4022 & -5260 & 3.78 & -968 & -354 \\
Na & 5.214 & -174.025 & 4.428 &  -837 &  -663 & 5.10 & -381 &  -27 \\
K  & 5.786 & -252.567 & 5.084 & -1274 &  -831 & 5.67 & -569 &  -52 \\
Rb & 6.208 & -221.399 & 5.596 &  -995 &  -513 & 6.13 & -483 &  -15 \\
Cs & 6.581 & -246.786 & 5.992 & -1139 &  -562 & 6.52 & -536 &  -32 \\
\end{tabular}
\end{ruledtabular}
\end{table}

The quartet trimers all have equilibrium interatomic distances (at
$D_{3h}$ geometries) that are substantially shorter than those of
the triplet dimers, by an amount that decreases steadily down the
series from 0.94 \AA\ in Li$_3$ to 0.59 \AA\ in Cs$_3$. The
three-atom  potentials are all correspondingly deeper than
pairwise sums of dimer potentials, by a factor that is more than 4
for Li$_3$ but is 1.3 to 1.5 for the heavier alkalis. 
The non-additive contributions $V_3$ to the interaction energies 
at the equilibrium geometries vary from 
approximately 120\% for Li to 50\% for Cs. 
These figures are much larger than for systems such as the rare gas
trimers, where the non-additive contributions are closer to 
0.5\% - 2.5\%  
\cite{raregas1,raregas2}
and produce a weakening rather than a strengthening of the binding.
However, the figures are quite similar to those for the 
alkaline-earth trimers, where the non-additive contributions range 
from about $100\%$ for Be$_{3}$ to 60\% for Ca$_{3}$ \cite{ale3}.

The RCCSD(T) two-body and three-body interaction potentials can be
decomposed into self-consistent field (SCF) and correlation
contributions.
The correlation contribution dominates at long range, but is
overcome by the SCF contribution when orbital overlap is
significant. For triplet alkali dimers, as for rare gas dimers,
the SCF potentials are repulsive and the main attractive forces
arise from interatomic correlation (dispersion).

The qualitative similarity between alkali and rare gas atoms does
not, however, extend to three-body forces. For rare gases, a large
part of the three-body energy comes from the non-additive
dispersion interaction. The leading long-range term in this is the
Axilrod-Teller-Muto (ATM) triple-dipole term \cite{ATM1,ATM2},
which dies off as $r^{-3}$ in each of the interatomic distances;
the ATM term is repulsive near equilateral configurations but
attractive near linear configurations. Simulations of rare gas
solids and liquids using accurate pair potentials and the ATM term
as the {\em only} non-additive contribution have proved remarkably
accurate \cite{Barker}.

For alkali atoms, by contrast, there is a large {\it attractive}
contribution 
to the three-body non-additive energy that exists
even at the SCF level. The SCF values for $V_3$ at the potential minima are
within 20\% of the coupled-cluster values 
($\sim110\%$ for Li, $\sim90\%$ for Na, $\sim100\%$ for 
K and Rb, and $\sim120\%$ for Cs).
This arises because, for alkali and alkaline-earth atoms, 
there are vacant $n$p orbitals that lie relatively close to the
$n$s orbitals. The $n$p orbitals can form bonding and antibonding
molecular orbitals (MOs) of the same symmetry ($a_1^\prime$ and
$e^\prime$) as those formed from the $n$s orbitals. The sets of
MOs of the same symmetry interact, lowering the energy of the
occupied MOs and contributing to bonding. In chemical terms, this
is essentially sp hybridization. 

\begin{table}[t]
\caption{Natural atomic orbital populations of spin-polarized alkali trimers 
and dimers at the corresponding global minima.} 
\label{table2}
\begin{ruledtabular}
\begin{tabular}{lrrrrr}
 & \multicolumn{3}{c}{Trimer} & \multicolumn{2}{c}{Dimer} \\
\cline{2-4} \cline{5-6}
 & $n$s & $n$p$_{\mbox{r}}$ & $n$p$_{\mbox{t}}$ & $n$s &  $n$p$_{z}$ \\
\tableline
Li & 0.743 & 0.046 & 0.197 & 0.992 & 0.005 \\
Na & 0.985 & 0.003 & 0.009 & 0.998 & 0.001 \\
K  & 0.949 & 0.011 & 0.034 & 0.995 & 0.003 \\
Rb & 0.975 & 0.006 & 0.014 & 0.996 & 0.003 \\
Cs & 0.947 & 0.012 & 0.030 & 0.995 & 0.003 \\
\end{tabular}
\end{ruledtabular}
\end{table}

To verify that this is indeed the mechanism, we have carried out natural
orbital population analyses \cite{npa} of the SCF wavefunctions for quartet
alkali trimers. The results are shown in Table \ref{table2}. At $D_{3h}$
geometries, the three atoms lie on a circle and the p-orbital populations
may be separated into radial and tangential parts. 
For Li$_3$, the population in radial $n$p$_{\rm r}$ orbitals is 0.046 and
that in tangential $n$p$_{\rm t}$ orbitals is 0.197. For the other alkalis
the fractions are much smaller, but still significantly larger than the
populations of $n$p$_{z}$ orbitals in the corresponding triplet dimers.

We conclude that non-additive terms make substantial contributions
to the potential energy surfaces for three-atom systems involving
spin-polarized alkali atoms. For Li, the full potential including
non-additive terms is a factor of 4 deeper than suggested by
pairwise additivity. For all the alkalis, the non-additive forces
are strongest at geometries near the equilateral
equilibrium configuration. A large part of the non-additive forces
exists even at the SCF level, and arises from sp mixing effects of a
type that cannot exist in rare gas systems. 
Non-additive dispersion forces are important at long range, but make a
relatively small contribution around the potential minimum.

We are currently working on generating complete potential energy 
surfaces for the alkali trimers, for use in quantum dynamical collision
calculations at ultralow energies.

\acknowledgments

The authors are grateful to the EPSRC for support under research
grant no.\ GR/R17522/01. JMH is grateful to JILA, University of
Colorado and National Institute of Standards and Technology, for
its hospitality in 2001-02.

\end{document}